\documentclass[12pt]{article}
\usepackage[dvips]{graphicx}
\newcommand{\simgt}{\hbox{ \raise3pt\hbox to 0pt{$>$}\raise-3pt\hbox{$\sim$} }}
\newcommand{\simlt}{\hbox{ \raise3pt\hbox to 0pt{$<$}\raise-3pt\hbox{$\sim$} }}
\newcommand{\DS}{\displaystyle }

\begin{document}
\begin{flushright}
\begin{tabular}{l}
OCHA-PP-221 \\
AJC-HEP-34 \\
April 30, 2004
\end{tabular}
\end{flushright}
\vfil
\begin{center}
{\LARGE
`TeV Gamma-ray Crisis' and \\[3MM]
an Anisotropic Space Model
}
\vfil \vfil
{Gi-Chol CHO}$^{\ref{1},\ref{2},}$%
\footnote{e-mail: {\tt cho@phys.ocha.ac.jp}}, 
{Jun-ichi KAMOSHITA}$^{\ref{1},\ref{2}}$, 
{Mariko MATSUNAGA}$^{\ref{1}}$ 
\footnote{Address after April 1, 2004: 
Communication Platform Laboratory, 
Corporate Research \& Development Center, 
Toshiba Corporation, 
1, Komukai Toshiba-cho, Saiwai-ku, Kawasaki  212-8582, Japan.
}
, \\[2mm]
{Akio SUGAMOTO}$^{\ref{1},\ref{2}}$ 
and 
{Isamu WATANABE}$^{\ref{3},}$%
\footnote{e-mail: {\tt isamu@akeihou-u.ac.jp}}
\\[1cm]
{\footnotesize 
\begin{enumerate}
\item
{\em Graduate School of Humanities and Sciences, Ochanomizu University, \\
     1-1 Otsuka 2-chome, Bunkyo, Tokyo 112-8610, JAPAN.} 
\label{1} \\[2mm]
\item
{\em Department of Physics, Ochanomizu University, \\
     1-1 Otsuka 2-chome, Bunkyo, Tokyo 112-8610, JAPAN.} 
\label{2} \\[2mm]
\item
{\em Department of Economics, Akita Keizaihoka University, \\
     46-1 Morisawa, Sakura, Shimokita-te, Akita 010-8515, JAPAN.} 
\label{3}
\end{enumerate}
}
\vfil \vfil
{\sc Abstract}
\end{center}
To solve the `TeV gamma crisis', we examine a model whose one spatial 
direction is discretized at a high energy scale.  
Assuming the standard extra-galactic IR photon distribution, we 
evaluate the mean free-path of a energetic photon which acquires an 
effective mass in the model.  
For a wide range of the value of the lattice energy scale 
between a few TeV and around $10^{10}$ GeV, the mean free-path of a TeV 
energy photon can be enlarged enough to solve the `crisis'.  
Taking into account the effects of the universe expansion, we find 
bounds of the lattice constant.

\newpage

\section{Introduction}

Gamma-rays of TeV energies from distant BL Lacertae objects, or `blazars', 
provide us a problem so-called `TeV gamma crisis' \cite{P&M}.  
For example, HEGRA collaboration observed the gamma-ray from Markarian 501 
(Mrk501) in 139 Mpc distance, and the maximum measured energy of the 
gamma-ray reaches to about 20 TeV \cite{M501}.  
An energetic gamma-ray can be attenuated by scatterings with the 
extra-galactic infrared photons in $\gamma\gamma \rightarrow e^+e^-$ process.  
Since the evaluated mean free-path of a 20 TeV photon is only 13 Mpc, 
the gamma-rays should have been suppressed by a factor of $\exp(-139/13)$ 
= $1/(4 \times 10^4)$.  
It means that the original brightness of Mrk501 in 20 TeV is $4 \times 10^4$ 
times brighter than what HEGRA saw.  
Comparing the observed energy spectrum of the gamma-rays of Mrk501 with 
that of much closer TeV gamma-ray sources like the Crab nebula, such a strong 
suppression in high energy side in the energy spectrum seems to be unrealistic.  
The same problem also holds for Mrk421, 1ES1959+650 and H1426+428 \cite{four}.  
The observed highest energies of the photons from these blazars are listed 
in \ref{T1}.  

Some trials to solve the `crisis' have been proposed by many authors.  
The most conservative way is to assume a less density of the extra-galactic 
infrared photons.  
One can also evaluate the appropriate IR photon density so that the 
traveling gamma-rays do not absorbed so much \cite{four}.  
There is another possibility to assume the violation of the Lorentz invariance 
in the space-time \cite{many,P&M}.  
Such kind of idea is motivated by a recent development of the theories of 
the large extra-dimension.  
In the deformed space-time, it is expected that the energy threshold of the 
$\gamma\gamma \rightarrow e^+e^-$ process can be shifted to reduce the 
absorption of the TeV gamma-rays.  

In this paper, we examine the possibility to answer the `TeV gamma crisis' 
by a model with reduced degree of space freedom at a high energy scale 
\cite{Suga}.  
In a model whose one spatial direction is discretized at a high energy 
scale, we evaluate the extra-galactic absorption of an energetic photon 
which acquires an effective mass in the model.  

The paper is organized as follows.  
In the section \ref{S2}, the adopted model is presented.  
The threshold shift of the scattering process and the mean free-path of 
the high energy photon in this model are evaluated in the section \ref{S3}.  
The effects of the universe expansion on the photon propagation is discussed 
in the section \ref{S4}.  
We give the conclusions in the last section.

\section{The Model}
\label{S2}

In the recent theories with large extra dimensions, we consider that the 
usual four dimensional space-time is located on a world volume of the brane 
(the extended object like membrane) and the world volumes stand parallel 
to each other with a separation distance (or lattice constant) of $a$.  
At high energy, such lattice structure becomes manifest itself and we are 
confined into the dynamics on a single brane, but after the universe cools 
down, the neighboring branes which are irrelevant so far, are possible to 
be dynamically connected and we arrive at the five dimensional theory at 
low energy.  
This was shown to happen in gauge theories in \cite{A-H}.
Following this idea, we are tempted to examine that the space-time 
dimensions become smaller and smaller according to the increase of energy 
scale in various theories as is discussed in \cite{Suga}.  

Here, in order to solve the `TeV gamma crisis', we look into the model in 
which the usual four dimensional QED works at low energy, but at high energy 
the lattice structure of one spatial dimension becomes manifest itself and we 
arrive at the three dimensional QED.  
More explicitly, we consider the theory in which the $x^3$ direction is 
discretized with the lattice constant $a$ of the space structure.  
See \ref{F0} for an illustration of the model.  

If the discrete dimension is written as $x^3$ $= n \cdot a$ ($n$:\ integer), 
the  Klein-Gordon equation for a boson field $\phi$ becomes 
\begin{equation}
\begin{array}{lcc}
\DS 
\left\{ (\frac{\partial}{\partial t})^2
       -(\frac{\partial}{\partial x^1})^2
       -(\frac{\partial}{\partial x^2})^2 \right\} \phi(t, x^1, x^2, n) 
 & & \\[2mm]
\DS 
-\frac{1}{a^2} 
\left\{    \phi(t, x^1, x^2, n+1) 
       - 2 \phi(t, x^1, x^2, n) 
         + \phi(t, x^1, x^2, n-1) \right\}
 & & \\[2mm]
-m^{*2} \phi(t, x^1, x^2, n) 
 & = & 0 , 
\end{array}
\end{equation}
where $t$, $x^1$ and $x^2$ are the continuum time and space coordinates, 
$m^*$ is the boson mass.  

In the momentum space it becomes 
\begin{equation}
\left( E_\gamma^2 - k_1^2 - k_2^2 - \tilde{k}_3^2 - m^{*2} \right) \phi = 0 , 
\end{equation}
where $E_\gamma$ and $(k_1,k_2,\tilde{k}_3)$ are the energy and the momentum 
of the boson with, 
\begin{equation}
\tilde{k}_3 = \frac{2}{a} \sin \frac{ak_3}{2} .  
\label{ktilde}
\end{equation}
Now we can understand that $k_3$ is replaced by $\tilde{k}_3$ for the discrete 
dimension.  
Such replacement is valid also for a photon.  

The fermion field $\psi$ describing electron satisfies the following Dirac 
equation:  
\begin{equation}
\begin{array}{lcc}
\DS 
\left\{ i\gamma^{0} \frac{\partial}{\partial t}
       +i\gamma^{1} \frac{\partial}{\partial x^1}
       +i\gamma^{2} \frac{\partial}{\partial x^2} \right\} \psi(t, x^1, x^2, n) 
 & & \\[2mm]
\DS 
+ i\gamma^{3} \frac{1}{2a} 
\left\{ \psi(t, x^1, x^2, n+1)-\psi(t, x^1, x^2, n-1) \right\} 
 & & \\[2mm]
-M \psi(t, x^1, x^2, n) 
 & = & 0 .  
\end{array}
\end{equation}
In the momentum space it becomes
\begin{equation}
\left( \gamma^{0}E_e 
      +\gamma^{1}p_1 
      +\gamma^{2}p_2 
      +\gamma^{3}\tilde{p}_3 
      -M \right) 
\psi = 0 ,
\end{equation}
where $E_e$ and $(p_1,p_2,\tilde{p}_3)$ are the energy and the momentum of 
the fermion with, 
\begin{equation}
\tilde{p}_3 = \frac{1}{a} \sin ap_3 .  
\label{ptilde}
\end{equation}

As for the interaction of electron and photon,
the gauge field (the connection field) $A_3$ in the $x^3$ direction is 
generated dynamically between the neighboring branes at $x^3=na$ and 
$(n+1)a$, giving
\begin{equation}
\bar{\psi}(t, x^1, x^2, n) \gamma_3 U(n, n+1) \psi(t, x^1, x^2, n+1) 
+ \mbox{h.c.}, 
\end{equation}
with 
\begin{equation}
U(n, n+1) = 
\exp \left[ i e a A_3(t, x^1, x^2, n) \right] 
= 1 + i e a A_3(t, x^1, x^2, n) + \cdots
\label{exp}
\end{equation}
In the expansion (\ref{exp}), the first term contributes to the Dirac 
equation discussed above, and the second ${\mit O}(a)$ term contributes 
to the interaction. 
The lowest interaction reads 
\begin{equation}
i e a \bar{\psi}(t, x^1, x^2, n) \gamma_3  A_3(t, x^1, x^2, n) 
           \psi (t, x^1, x^2, n+1) + \mbox{h.c.}.  
\end{equation}
The incoming fermion and the outgoing fermion are located on the different 
branes, so that the gauge interaction is modified from the usual QED by the 
following extra factor: 
\begin{equation}
\frac{1}{2} \left( e^{-iap_3} + e^{aip'_3} \right) ,
\end{equation}
where $p'_3$ and $p_3$ are, respectively, the third component of the incoming 
and outgoing fermion momenta.
However, such modification of interactions from the standard QED always 
associate the factor ((lattice constant $a$) $\times$ (momentum or 
energy))$^2$, where the square guarantees the parity conservation.  
Therefore, the modification of the interaction is small at energy lower 
than $a^{-1}$.  
It is also expected that the QED in our model has additional vertices 
of photons with electrons.  
However, the effect of such deformation of the QED is expected to be less 
important for the `crisis' problem than that of the shift of the threshold.  
The most important change of the QED for the `crisis' is the energy-momentum 
relation which provides the threshold shift of the $\gamma\gamma \rightarrow 
e^+e^-$ process, as will be presented below in the section \ref{S3}.  
In this paper, we concentrate on the effect of the change of the 
energy-momentum relation.  

In a way of thinking, particles are assumed to have always the usual 
energy-momentum relations such as, 
\begin{equation}
E_\gamma^2 - k_1^2 - k_2^2 - k_3^2 = 0 .  
\label{Eg0}
\end{equation}
But at high energy scale, the $x^3$ space is curled up to a circle with 
the circumference $a$.  
Then, in this deformed space, the equation of motion is written using 
$\tilde{k}_3$ instead of $k_3$ in equation (\ref{Eg0}).  
Therefore, the energy-momentum relation can be viewed effectively as 
\begin{equation}
E_\gamma^2 - k_1^2 - k_2^2 - \tilde{k}_3^2 = m^{*2} , 
\label{Eg} 
\end{equation}
having an effective mass 
\begin{equation}
m^{*2} = 
k_3^2 \left[ 1 - \left( \frac{2}{ak_3} \sin \frac{ak_3}{2} \right)^2 \right] .  
\label{m*}
\end{equation}
A similar relationship also holds for a fermion with the intrinsic mass $m_e$:  
\begin{eqnarray}
E_e^2 - p_1^2 - p_2^2 - p_3^2 & = & m_e^2 , 
\label{Ee0} \\
E_e^2      - p_1^2 - p_2^2 - \tilde{p}_3^2 & = & M_e^2 
= m_e^2 + m_e^{*2} , 
\label{Ee} \\
m_e^{*2} & = & 
p_3^2 \left[ 1 - \left( \frac{1}{ap_3} \sin ap_3 \right)^2 \right] .  
\label{me*}
\end{eqnarray}
It is essential that the additional component of the effective mass $m_e^*$ 
only depends on $a$ and $p_3$, but does not depend on $m_e$.  
Expanding the equations (\ref{m*}) and (\ref{me*}) in powers of $ak_3$ and 
$ap_3$, respectively, one gets, 
\begin{eqnarray}
m^{*2}   & = & k_3^2 \left[ \frac{1}{12} (ak_3)^2 
                          + \frac{1}{360} (ak_3)^4 + \cdots \right] , 
\label{mm*} \\
m_e^{*2} & = & p_3^2 \left[ \frac{1}{3} (ap_3)^2 
                          + \frac{2}{45} (ap_3)^4 + \cdots \right] .  
\label{mme*}
\end{eqnarray}

\section{Threshold Effect to Mean Free-Path}
\label{S3}

The scattering process of the blazar gamma-rays $\gamma\gamma \rightarrow e^+e^-$ 
is characterized by an asymmetric initial photon energies; {\it i.e.}, 
the energy of the traveling photon $E_\gamma$ is in the order of TeV, 
while that of the extra-galactic IR photon $\epsilon$ is in the order of eV.  
Thus, the effect of the deformation of the space structure to the extra-galactic 
IR photon is totally negligible.  
The energies of the produced electron $E_{e1}$ and positron $E_{e2}$ are also 
in the order of TeV in general.  

Especially, at the threshold, the electron and the positron have the same 
energy $E_e$ $= E_{e1}$ $= E_{e2}$ $\simeq E_\gamma/2$.  
Assuming the blazar photon travels along to the $x^3$ direction, we can 
write down the energy-momentum conservation law just on the threshold as 
follows:  
\begin{eqnarray}
E_\gamma              + \epsilon & = & 2 E_e         , \\
E_\gamma \beta_\gamma - \epsilon & = & 2 E_e \beta_e , 
\end{eqnarray}
where $\beta_\gamma$ $= \sqrt{1-m^{*2}/E_\gamma^2}$ and $\beta_e$ 
$= \sqrt{1-M^2/E_e^2}$ are the velocities of the energetic photon and the 
electron in the laboratory frame, respectively.  
Since $E_\gamma$ $\simeq k_3$ and $E_e$ $\simeq p_3$, $m^*$ $\simeq 2 m_e^*$ 
at the threshold.  
In such a case, $M_e$ is dominated by $m_e^*$ for, 
\begin{equation}
a^{-1} < E_\gamma^2/(4\sqrt{3}m_e) , 
\label{thres}
\end{equation}
which results in a significant increase of the threshold energy and a desirable 
suppression of the absorption process.  
For $E_\gamma$ = 20 TeV, $a^{-1}$ $< 1 \times 10^{11}$ GeV which means that 
the deformation of the space structure at a very high energy scale can affect 
the electron-positron pair production at the scale of MeV.  

The mean free-path of the energetic photon $x_{\gamma\gamma}$ can be 
evaluated by the equation (1) in the reference \cite{P&M}:  
\begin{equation}
x_{\gamma\gamma}^{-1} = 
\frac{1}{8E_\gamma^2\beta_\gamma} 
\int^{\infty}_{\epsilon_{\mbox{\tiny min}}} 
d\epsilon \frac{n(\epsilon)}{\epsilon^2} 
\int^{s_{\mbox{\tiny max}}}_{s_{\mbox{\tiny min}}} 
ds ( s - m^{*2} ) \sigma , 
\label{mean}
\end{equation}
where $n(\epsilon)$ is the number density distribution of the extra-galactic 
IR photon of the energy between $\epsilon$ and $\epsilon+d\epsilon$ which 
can also be found in the figure 1 in the reference \cite{P&M}, $s$ 
$= 2 E_\gamma \epsilon (1+\beta_\gamma) + m^{*2}$ is the C.M.\ energy 
squared of the process, and, 
\begin{eqnarray}
\epsilon_{\mbox{\tiny min}} & = & 
\frac{4 M_e^2 - m^{*2}}{2 E_\gamma ( 1 + \overline{\beta}_\gamma )} 
\simeq 
\frac{4 m_e^2}{2 E_\gamma ( 1 + \overline{\beta}_\gamma )} , \\
s_{\mbox{\tiny max}} & = & 
m^{*2} + 2 \epsilon E_\gamma ( 1 + \overline{\beta}_\gamma ), \quad 
s_{\mbox{\tiny min}} = 4 M_e^2 , 
\end{eqnarray}
where $\overline{\beta}_\gamma$ $= 1 - m^{*2}/s$ is the average initial 
photon velocity in the C.M.\ frame.  
We used for simplicity the formula of the total cross section of the 
scattering process $\sigma$ as same as in the ordinary QED, which is 
described as, 
\begin{equation}
\sigma = \frac{4\pi\alpha^2}{s\overline{\beta}_\gamma} 
\left[ \frac{3-\overline{\beta}_e^4}{2} 
       \ln \frac{1+\overline{\beta}_e}{1-\overline{\beta}_e} 
       - \overline{\beta}_e \left( 2 - \overline{\beta}_e^2 \right) \right] , 
\label{sigma}
\end{equation}
where $\alpha$ the QED fine structure constant, $\overline{\beta}_e$ 
$= \sqrt{1 - 4M_e^2/s}$ the electron velocity in the C.M.\ frame, 
respectively.  
We adopted an approximation such that the effective masses of the electron 
and the positron has the same value $m_e^*$ $= m^*/2$ as just as at the 
threshold.  
We took into account the effective mass effect of the photon only in the 
flux factor in the denominator of the first fraction in the r.h.s.\ of 
the formula (\ref{sigma}).  
The cross section $\sigma$ is illustrated in \ref{Fs} for the ordinary 
space-time with 0.511 MeV electron mass.  
The peak of the cross section can be found at 1.40 times above the 
threshold.  

\ref{F1} shows the evaluated value of the mean free-path of the traveling 
photon $x_{\gamma\gamma}$ as a function of the photon energy $E_\gamma$ 
and the lattice constant of the space structure $a$.  
The lowest curve of $a^{-1}$ $= \infty$ case corresponds to the ordinary 
continuum space-time.  
Here the upper edges of the curves, which corresponds the lower bounds 
of $a^{-1}$ for a fixed $E_\gamma$, can be derived from a restriction of 
$aE_\gamma/2$ $< \pi$, which proves the photon propagates to the correct 
direction.  

We found that the mean free-path is enlarged drastically for a wide range 
of the value of $a^{-1}$.  
One can observe from the figure, departure of the curves from the $a^{-1}$ 
$= \infty$ case happens around $a^{-1}$ $= E_\gamma^2/(4\sqrt{3}m_e)$, 
as is already suggested in the equation (\ref{thres}).  
The mean free-path in the modified space structure is characterized by 
the followings:  
\renewcommand{\theenumi}{(\roman{enumi})}
\begin{enumerate}
\item 
A minimum of the mean free-path is found at $E_\gamma$ around several 
hundred GeV to 1 TeV for $a^{-1}$ $< 10^9$ GeV.  
\label{c1}
\item 
Almost constant value of the mean free-path is found at $E_\gamma$ around 
10 TeV and $a^{-1}$ $\simlt 10^{10}$ GeV, which is in contrast with the 
$a^{-1}$ $= \infty$ case.  
It means that there is no energy cut-off of the blazar photon spectrum 
around 10 TeV.  
\label{c2}
\item 
For a large $E_\gamma$, the mean free-path is raised dramatically as is 
proportional to $a \cdot E_\gamma^{5/2}$.  
\label{c3}
\end{enumerate}

The \ref{F2} shows that a 20 TeV photon has a longer mean free-path 
than 139 Mpc of Mrk501 distance for $10^{3.5}$ GeV $\simlt a^{-1}$ 
$\simlt 10^{10}$ GeV.

\section{Photon Propagation in the Expanding Universe}
\label{S4}

More accurately, one can take into account the effect of the universe 
expansion during the photon propagation from Mrk501 to the earth.  
This effect is only important near the lower bound of $a^{-1}$ which 
results in a larger value of $m^*$ and slower $\beta_\gamma$.  

We assumed a de Sitter expanding universe which is spatially 
uniform at the temperature below $a^{-1}$.  
The scale factor $R(t)$ of the universe has a time $t$ dependence as, 
\begin{equation}
\frac{d R(t)}{dt} = H_0 \sqrt{\Omega_m/R + \Omega_v R^2} , 
\label{expand}
\end{equation}
where $H_0$ the Hubble constant, $\Omega_m$ the present matter density 
fraction and $\Omega_v$ the present vacuum energy density fraction, 
respectively.  
We adopted $H_0$ = 71 km/s/Mpc, $\Omega_m$ = 0.27 and $\Omega_v$ = 0.73, 
according to the recent results by WMAP \cite{WMAP}.  
During the universe expansion, the energy of the traveling photon 
decreases due to the red-shift.  
We assumed that the evolution of the number density distribution of 
the extra-galactic IR photon is only due to the cosmological expansion.  
The lattice constant $a$ is assumed to be constant during the time 
of interest of the photon propagation.  

We can evaluate the distances of blazar galaxies by the Hubble law, 
neglecting any proper motions in the co-moving frame of the universe.  
At the time when the visual light we see today on the earth was departed 
from the blazar at $t_*$, the scale factor of the universe expansion was, 
\begin{equation}
R(t_*) = R_0 / (1+z) , 
\end{equation}
where $z$ is the red-shift of the observed light in the visual wavelength, 
$R_0$ is the present scale factor.  
The present distance of the blazar in the co-moving frame $y_*$ can be 
evaluated as, 
\begin{equation}
y_* = \int_{t_*}^0 dt \frac{R_0}{R(t)} , 
\end{equation}
where we set the value of the present time to be zero.  
Since an energetic photon with an effective mass travels at a slower 
speed $\beta_\gamma$ than unity, such a photon should depart at 
an earlier time $\tilde{t}_*$ than $t_*$ of the ordinary light:  
\begin{equation}
y_* = \int_{\tilde{t}_*}^0 dt \frac{R_0}{R(t)} \beta_\gamma .  
\end{equation}
Note that $\beta_\gamma$ is a function of the scale factor $R(t)$, 
then it depends on $t$.  
The degree of the transparency $T$ of an energetic photon can be 
evaluated as, 
\begin{equation}
T = \exp \left[ 
   - \int_{\tilde{t}_*}^{0} dt \frac{\beta_\gamma}{x_{\gamma\gamma}}
         \right] , 
\end{equation}
here $x_{\gamma\gamma}$ is also a function of $R(t)$.  

\ref{F2} displays the transparency $T$ dependences on the lattice 
constant $a$, for a 20 TeV photon from Mrk501 ($z$ = 0.0336) and 
a 5.37 TeV photon from H1426+428 ($z$ = 0.129).  

For Mrk501, we found $10^{3.4}$ GeV = 2.5 TeV $< a^{-1}$ 
$\simlt 10^{10.5}$ GeV.  
Here we set the upper bound of $a^{-1}$ so that the transparency $T$ 
should be at least around 1/10.  
To reach a 20 TeV photon from Mrk501 to the earth, $aE_\gamma(t)/2$ 
should be greater than $\pi$ even at the time of departure when the 
photon energy $E_\gamma$ has its maximum value.  
The lower bound can be derived from this condition.  

One should recall here that we have assumed that the energetic photon 
propagates to the third direction which is in the discretization.  
If the photon travels in the direction with the polar angle $\theta$, 
the third component of the momentum has the fraction of $\cos\theta$ 
of the absolute value of the momentum, and thus the limits of $a^{-1}$ 
described above should be multiplied by $\cos\theta$ in this case.  

The same argument can be repeated to the other blazars listed in 
\ref{T1}.  
Among them, a new bounds to $a^{-1}$ can be derived from the 5.37 TeV 
photons from H1426+428, where we found $10^{3.14}$ GeV = 1.4 TeV 
$< a^{-1}$ $\simlt 10^{9.5}$ GeV.  
Note here that the angle between Mrk501 and 1H1426+428 is 27$^\circ$, 
so the limits of two blazars cannot be combined into a single 
expression trivially.

\section{Conclusions}
\label{S5}

To solve the `TeV gamma crisis', we have examined a model whose one 
spatial direction is discretized at a high energy scale.  
Assuming the standard extra-galactic IR photon distribution, we have 
evaluated the mean free-path of an energetic photon which acquires an 
effective mass in the model.  
For a wide range of the value of the lattice energy scale $a^{-1}$ 
between a few TeV and around $10^{10}$ GeV, the mean free-path of a TeV 
energy photon can be enlarged enough to solve the `crisis'.  
Taking into account the effects of the universe expansion, we have found 
bounds of $a^{-1}$: 2.5 TeV $< a^{-1}$ $\simlt 10^{10.5}$ GeV for 
a 20 TeV photon from Mrk501, and 1.4 TeV $< a^{-1}$ $\simlt 10^{9.5}$ 
GeV for a 5.37 TeV photon from H1426+428, under an assumption that the 
blazar gamma-ray travels to the discretized direction.  

Further test of the present model can be performed by watching PeV 
photons.  
In the ordinary space-time, a 1 PeV photon has the mean free-path of  
9 kpc.  
However, it can travel over 300 Mpc in our model with $a^{-1}$ 
$\simlt 10^{10}$ GeV, as can be seen in \ref{F1}.  
Observation of PeV photons from the near-by galaxies will support our 
model, which is a consequence of the item \ref{c3} described in the 
end at the section \ref{S3}.  
For $a^{-1}$ $\sim 10^8$--$10^9$ GeV, a 1 TeV photon has shorter 
mean free-path than a few to several TeV photons in our model, as was 
also mentioned in the item \ref{c1}.  
It results in a dent of the photon energy spectrum around 1 TeV for 
the objects which are distant as hundred Mpc.  

We also mention some other possible effect of the space discretization.  
In the case of smaller values of $a^{-1}$ as a few TeV to several ten 
TeV, collider experiments may find some direct evidence that the 
energetic light particles fly slowly due to the effective mass.  
Modification to the QED vertex structure may affect to the magnetic 
and the electric dipole moments of the elementary charged particles, 
even in the interactions with low energy photons.  
These observations will bring us additional bounds of $a^{-1}$.

\section*{Acknowledgment}
The authors were supported by a Grant-in-Aid for Scientific Research 
of No.\ 14039204 from the Ministry of Education, Culture, Sports, 
Science and Technology, Japan.
One of the authors, G.C.C.\ is also supported by a Grant-in-Aid 
for Scientific Research of No.\ 15740146.

\newpage
\section*{Table}

\renewcommand{\theenumi}{Table \arabic{enumi}}
\begin{enumerate}

\item 
The maximum energy of the gamma-rays from the blazars observed by HEGRA.  
We adopt the highest energy bin in the energy spectrum such that the 
differential energy distribution is measured well above its error.  
\begin{center}
\begin{tabular}{|l|l|r|c|c|}
\hline 
object & red-shift & \multicolumn{1}{l|}{maximum}  & significance & reference \\
name   &           & photon energy                 &              &           \\
\hline 
Mrk501      & 0.0336 & 18.45 TeV & 3.3$\sigma$ & \cite{M501} \\
Mrk421      & 0.031  & \multicolumn{2}{l|}{13 TeV (read from Figure)} 
                                               & \cite{M421} \\
1ES1959+650 & 0.047  & 10.94 TeV & 4.0$\sigma$ & \cite{1959} \\
H1426+428   & 0.129  &  5.37 TeV & 4.0$\sigma$ & \cite{1426} \\
\hline
\end{tabular}
\end{center}
\label{T1}

\end{enumerate}

\vfil

\section*{Figures}

\renewcommand{\theenumi}{Fig.\ \arabic{enumi}}
\begin{enumerate}

\item 
An illustration of our model with a discretized space direction $x_3$.  
The size of the lattice spacing $a$ is common for all gaps, and each 
membrane is numbered by an integer.  
\label{F0}

\item 
The collision energy $\sqrt{s}$ dependence of the total cross section of 
the process $\gamma\gamma \rightarrow e^+e^-$ in the ordinary space-time.  
The peak of the cross section can be found at $\sqrt{s}$ = 1.43 MeV = 1.40 
$\times (2m_e)$.   
\label{Fs}

\item 
The mean free-path $x_{\gamma\gamma}$ of the energetic photon as a function 
of the photon energy $E_\gamma$ and the lattice constant of the space 
structure energy scale $a^{-1}$.  
The lowest bold curve of $a^{-1}$ $= \infty$ corresponds to the ordinary 
continuum space-time.  
\label{F1}

\item 
The transparency $T$ dependences on the energy scale of the lattice constant 
$a^{-1}$, for a 20 TeV photon from Mrk501 (solid curve) and a 5.37 TeV photon 
from H1426+428 (dashed curve).  
The lower bounds of $a^{-1}$ are plotted by a square (Mrk501) and a cross 
(H1426+428).  
We assumed, for each case, that the blazar is on the third direction of 
discretization.  
\label{F2}

\end{enumerate}

\begin{figure}[p]
\begin{center}
\hspace*{-40mm}
\rotatebox{90}{\includegraphics{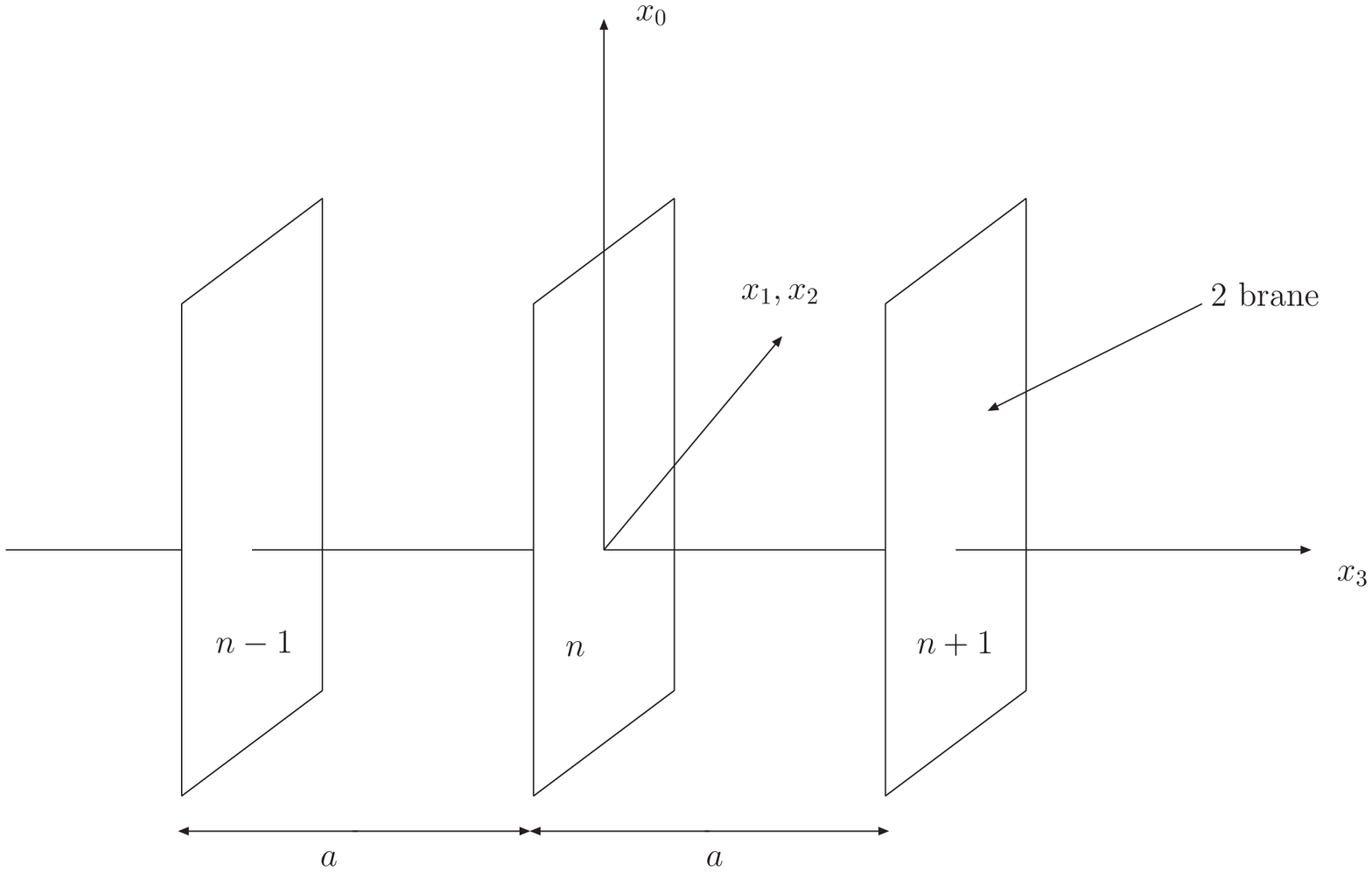}}
\\[1cm]
\ref{F0}
\end{center}
\end{figure}

\begin{figure}[p]
\begin{center}
\hspace*{-30mm}
\includegraphics{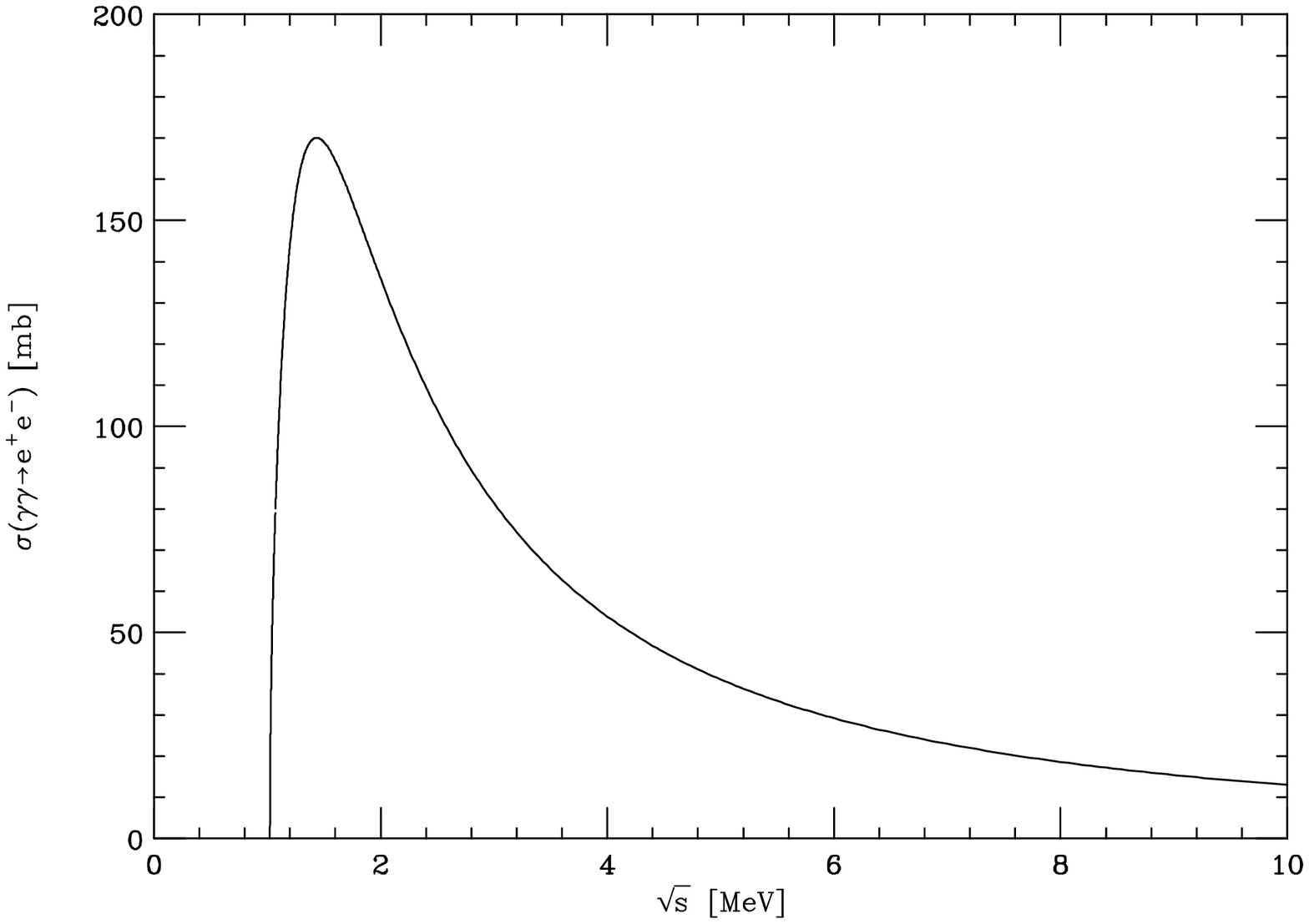}
\\[1cm]
\ref{Fs}
\end{center}
\end{figure}

\begin{figure}[p]
\begin{center}
\hspace*{-30mm}
\includegraphics{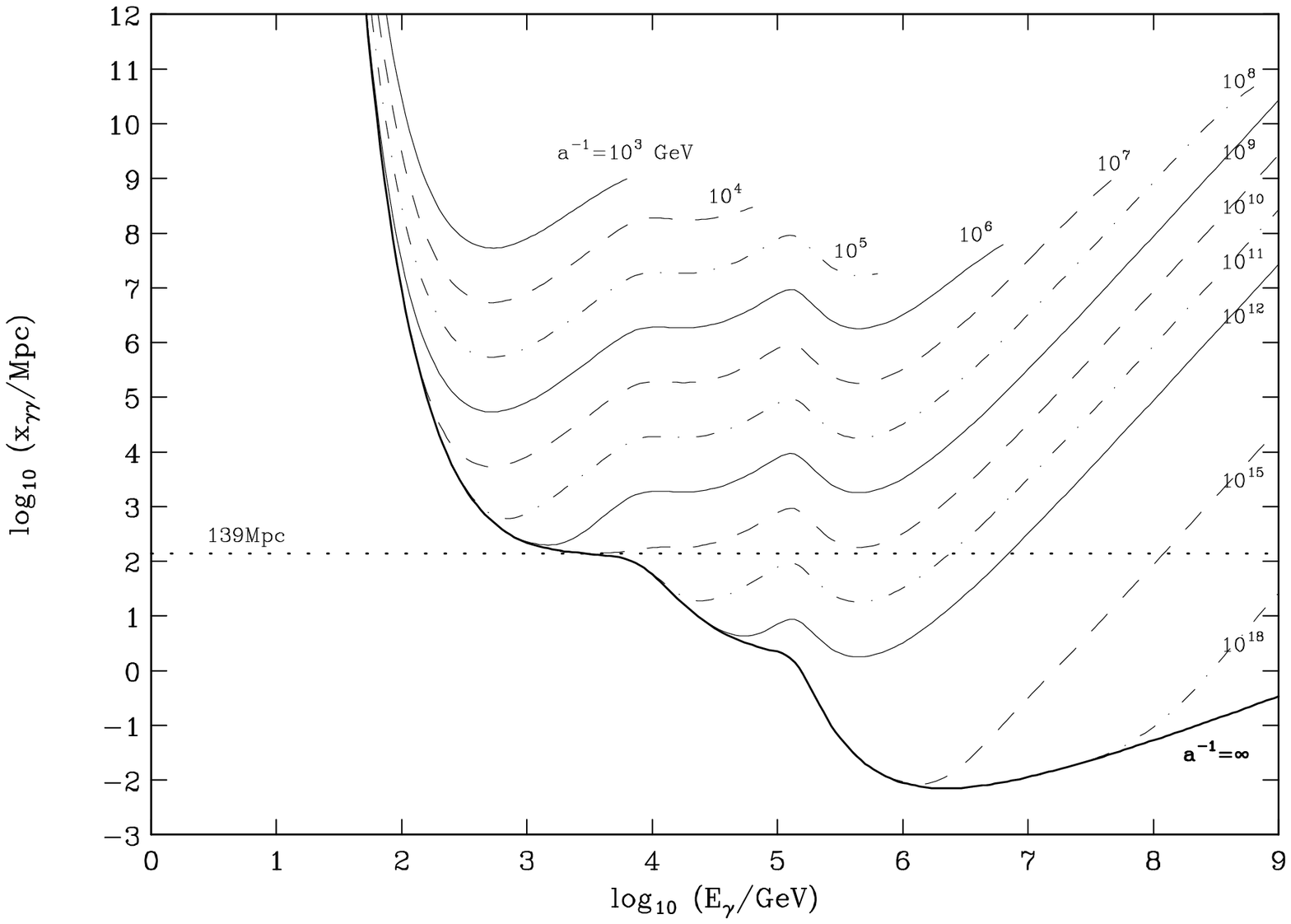}
\\[1cm]
\ref{F1}
\end{center}
\end{figure}

\begin{figure}[p]
\begin{center}
\hspace*{-30mm}
\includegraphics{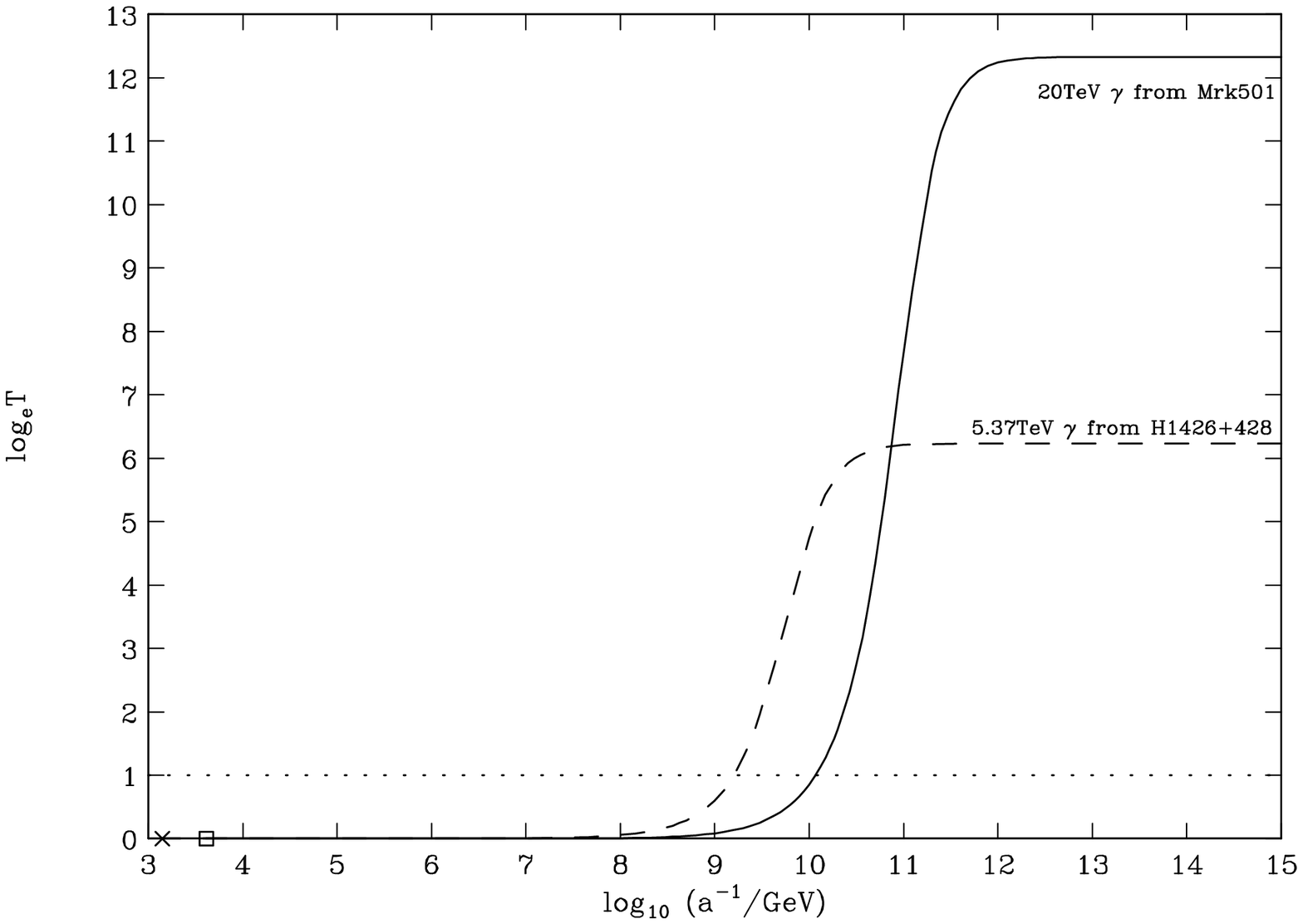}
\\[1cm]
\ref{F2}
\end{center}
\end{figure}

\end{document}